\documentclass[preprints,article,accept,moreauthors,pdftex]{mdpi} 

\firstpage{1} 
\pubvolume{xx}
\issuenum{1}
\articlenumber{5}
\pubyear{2019}
\copyrightyear{2019}
\history{} 

\pdfoutput=1

\usepackage{tabularx}
\usepackage{colortbl}
\definecolor{Gray}{gray}{0.9}
\definecolor{grannysmithapple}{rgb}{0.66, 0.89, 0.63}
\definecolor{lightcoral}{rgb}{0.94, 0.5, 0.5}
\definecolor{lightsalmonpink}{rgb}{1.0, 0.6, 0.6}
\definecolor{ashgrey}{rgb}{0.7, 0.75, 0.71}

\Title{BlackWatch: Increasing Attack Awareness Within Web Applications}

\Author{Calum C. Hall $^{1}$, Lynsay A. Shepherd $^{2}$\orcidA{}* and Natalie Coull $^{2}$}

\AuthorNames{Calum C. Hall, Lynsay A. Shepherd, Natalie Coull}

\address{%
$^{1}$ \quad MWR InfoSecurity, London, SE1 3RS, United Kingdom; calumhall96@gmail.com\\
$^{2}$ \quad School of Design and Informatics, Abertay University, Dundee, DD1 1HG, United Kingdom; lynsay.shepherd@abertay.ac.uk, n.coull@abertay.ac.uk}

\corres{Correspondence: lynsay.shepherd@abertay.ac.uk; Tel.: (+44) 01382308685}

\abstract{Web applications are relied upon by many for the services they provide. It is essential that applications implement appropriate security measures to prevent security incidents. Currently, web applications focus resources towards the preventative side of security.  Whilst prevention is an essential part of the security process, developers must also implement a level of attack awareness into their web applications. Being able to detect when an attack is occurring provides applications with the ability to execute responses against malicious users in an attempt to slow down or deter their attacks.  This research seeks to improve web application security by identifying malicious behaviour from within the context of web applications using our tool BlackWatch.  The tool is a Python-based application which analyses suspicious events occurring within client web applications, with the objective of identifying malicious patterns of behaviour. Based on the results from a preliminary study, BlackWatch was effective at detecting attacks from both authenticated, and unauthenticated users.  Furthermore, user tests with developers indicated BlackWatch was user friendly, and was easy to integrate into existing applications.  Future work seeks to develop the BlackWatch solution further for public release.}


\begin{document}

\section{Introduction}\label{intro}
Society is becoming increasingly dependent on the online services provided by many organizations \cite{riek2016} \cite{kumar2017study}. Web applications enable users to interact with organizations using their own portable devices. However, with the continued growth of online services comes the burden of ensuring appropriate security is implemented. Web applications are often the most public facing component within many organizations, and are therefore the typical attack vector chosen by malicious attackers \cite{rodriguez2012growing}. In Verizon’s 2018 Data Breach Incident Report, it was stated that approximately 19.5\% of data breaches analyzed by the Verizon research team were a result of web application attacks \cite{Verizon2018}.

Developers often focus their resources towards the preventative side of application security, to reduce incidents of web application attacks. However, work presented in this paper aims to demonstrate that prevention methods alone are not enough to protect an application- it is essential for developers to implement appropriate methods to detect that attacks are occurring against their applications. When targeting an application, very rarely will an attacker discover a vulnerability on their first attempt. Thus, if applications are able to accurately identify the behaviour of malicious users in a timely manner, then it becomes possible to execute the necessary responses required to deter or slow down these application attacks.  This in turn protects the web application itself, and protects innocent users from engaging with a vulnerable web application.

Following an overview of existing methods used for web application security, we present our solution, BlackWatch (named after the infantry battalion of the Royal Regiment of Scotland).  BlackWatch aims to identify malicious behaviour from within the context of a web application, and can be used by developers to increase the attack awareness.

\begin{figure}[ht]
\centering
\includegraphics[width=98mm]{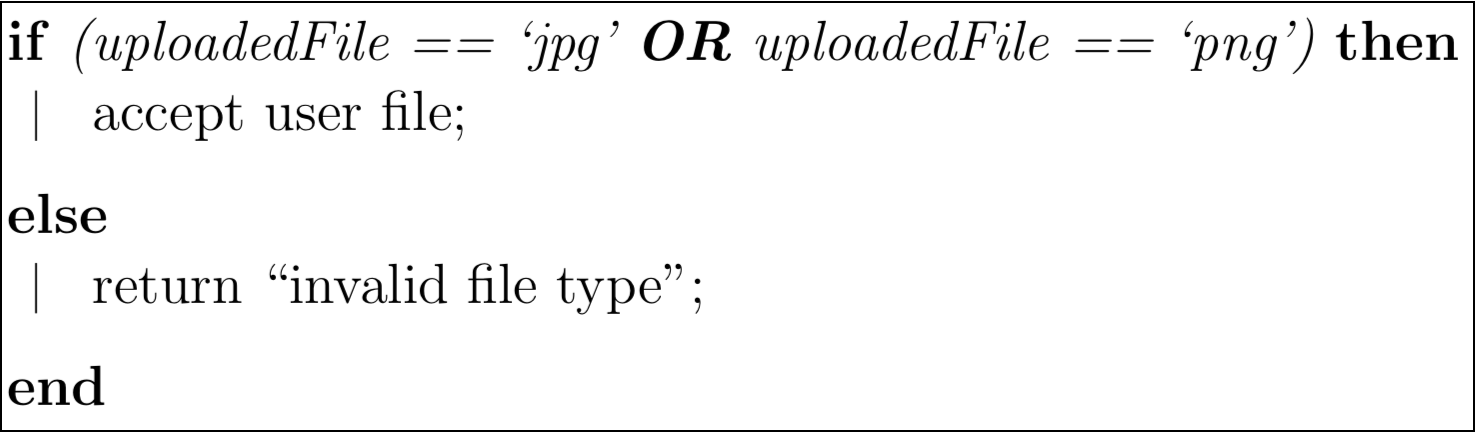}
\caption{File Upload}
\label{fig:fileupload}
\end{figure}

A key feature of BlackWatch is that the method of attack detection is designed to work within the web applications themselves. When creating an application, developers have a clear understanding of the legitimate usage that is expected at each stage, and hence they are able to accurately identify when a user’s actions are ‘unexpected’ or ‘suspicious’. An example scenario involves a web application containing a feature allowing users to upload a profile picture. The application would most likely limit the type of files users can upload through code similar to that shown in Figure \ref{fig:fileupload}.  
 
\begin{figure}[ht]
\centering
\includegraphics[width=98mm]{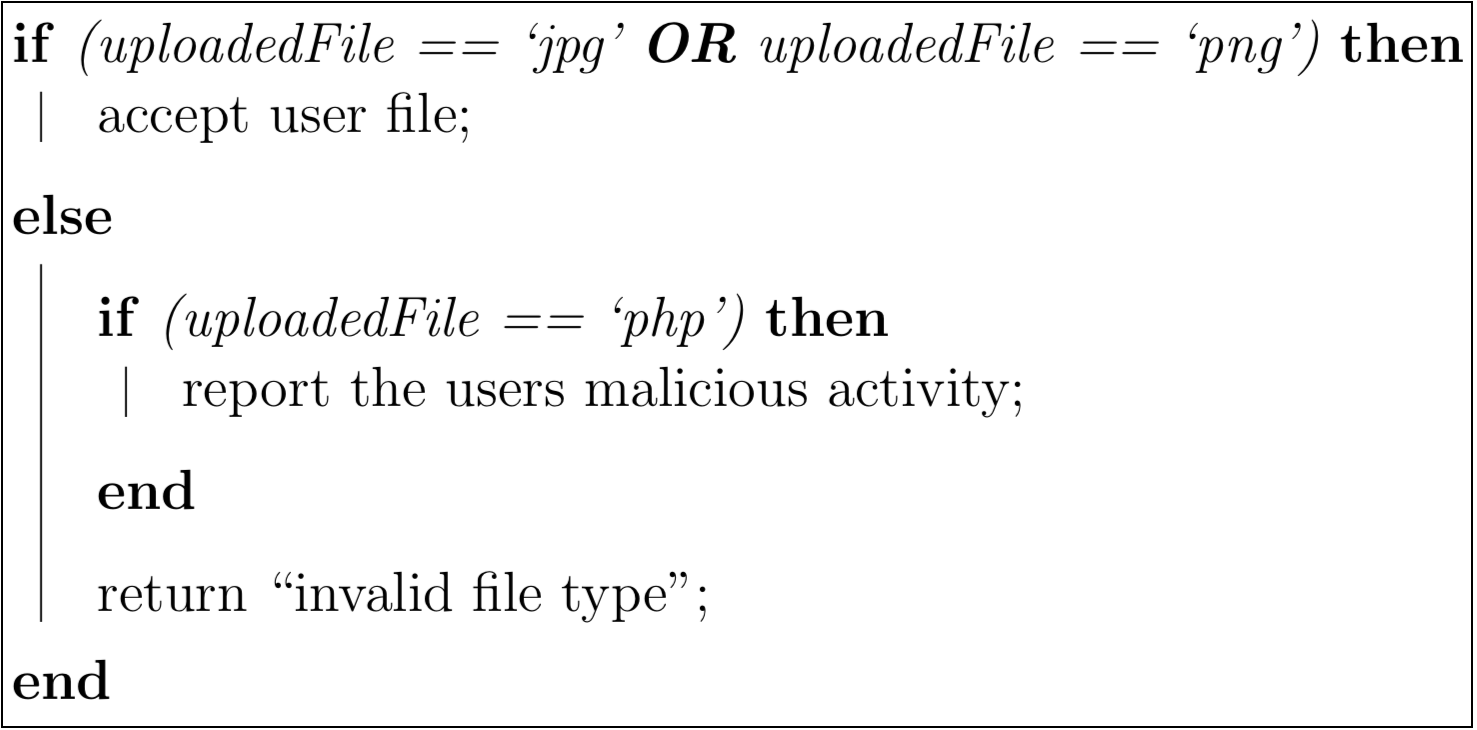}
\caption{File Upload Check}
\label{fig:fileuploadcheck}
\end{figure}

This preexisting function could be easily altered by a developer to implement further checks. For example, the application could check to see if the uploaded file is a PHP file, and if so it is highly likely that this is an attacker testing for a ‘malicious file upload’ vulnerability (Figure \ref{fig:fileuploadcheck}).  

The BlackWatch solution aims to monitor this type of suspicious behaviour, with the objective of analyzing user events to determine if their actions are indicative of malicious intent. 

This paper seeks to discuss the current security measures available for web applications in relation to firewalls in section \ref{background}.  The design and development of the BlackWatch solution will be covered in section \ref{methodology}, before evaluating the tool and presenting the results in section \ref{results}.  Finally, the results will be discussed in section \ref{discussion}, and conclusions are drawn in section \ref{conclusions}.

\section{Background}\label{background}
Web applications are a highly targeted component of many organizations, therefore it is essential that the security industry is actively improving methods used to defend against attacks. Insecure web applications are detrimental to online businesses, and can place end-users at risk of interacting with a vulnerable site \cite{shepherd2013perception}.  The section demonstrates the current state of web application development, highlighting how alternative methods may improve attack awareness.

\subsection{Attack Awareness}
In recent years, the security industry has switched focus from attack prevention to attack detection.  Whilst prevention is a vital aspect of security, detection must be taken into consideration.  The switch in focus has been discussed by industry experts: to ensure an organization has a strong level of security in place, they must move forward with the assumption that their prevention methods will be bypassed \cite{BenMeir2017}.  This approach allows organizations to focus their efforts on being security aware, to the extent where they can detect attacks in a timely manner, therefore allowing them to respond effectively. Recently it has been stated that, \textit{``security focus has shifted from prevention toward resilience and a broader set of capabilities including timely detection and the ability to respond to live incidents"} \cite{MWRInfoSecurity2018}.

\subsection{Current Industry Methods}
Within  industry a variety of methods and tools  are recommended for detecting and preventing attacks against web applications. The Payment Card Industry Data Security Standards suggests two alternative solutions, one of which must be implemented to ensure compliance regulations are met \cite{PCISSC2016}:
\begin{itemize}
\item Regular source code and component reviews
\item Detection and prevention of web-based attacks via the use of an automated solution, i.e. a web application firewall
\end{itemize}

Due to the extensive resources required for source code and component reviews, the majority of organizations choose to implement a web application firewall (WAF) instead. WAF solutions have the ability to effectively protect web applications from a large number of attacks, however to ensure effectiveness, organizations must invest a large quantity of resources into configuring and maintaining the solution. High-Tech Bridge \cite{Bridge2016} identify the issue by stating \textit{``ModSecurity is a very powerful and flexible WAF, however it requires a lot of time and effort to avoid false-negatives and prevent false-positives"}. Whilst High-Tech Bridge \cite{Bridge2016} are very positive about the usage of web application firewalls, not all security researchers express the same opinion. Kolochenko published an article exploring the five main downfalls facing web application firewalls \cite{Kolochenko2016}:

\begin{itemize}
\item \textit{``Negligent deployment, lack of skills and different risk mitigation priorities"}- There is often a lack of accountability within organizations as to the
entity responsible for maintaining the WAF.

\item \textit{``Deployment only for compliance purposes"}- Due to the PCI-DSS compliance regulations, WAF solutions are often deployed without ever being properly configured or managed.

\item \textit{``Complicated diversity of constantly evolving web applications"}- Web applications are constantly evolving and implementing new features, therefore keeping web application firewall rules up-to-date can be a challenging and time-consuming task.

\item \textit{``Business priorities domination over cyber security"}- A key concern of any security solution is the risk of affecting legitimate application usage.  To prevent false-positives from affecting user, organizations switch WAFs to `detect only' mode.  This means no preventive action is taken.  Further to this, many WAFs feature poor reporting capabilities: unless an employee is tasked with manually checking the reported incidents, these attacks often go unnoticed.

\item \textit{``Inability to protect against advanced web attacks"}- WAFs often work by blocking known attacks and identifying events that are predetermined as malicious; therefore new, advanced attacks will not be detected.

\end{itemize}

It is apparent that the most common downfall of WAF solutions is that often they are not configured properly, and default configurations are not effective. Security researcher Mazin Ahmed \cite{ahmed_2015} published work into web application firewalls where he demonstrated why default configurations are not effective. The approach of `blacklisting' that many web application firewalls use consists of defining a set of data that represents known-bad attacks \cite{clincy2018}. The issue with this type of approach as demonstrated by Ahmed \cite{ahmed_2015} is that this approach to security can be easily bypassed. Within this research, eight well renowned solutions were tested using their default configurations, and by using a variety of different methods Ahmed managed to bypass all eight solutions.

\subsection{Implementing Security into the Software Development Lifecycle (SDLC)}\label{sdlc}
One of the most difficult realities faced within the security industry, is that for an application to be secure, developers must consider every attack vector. However, for an application to be considered `insecure' an attacker only needs to find one area that hasn't been fully secured.  In an effort to combat this, there has been a concerted effort to encourage developers to include security into the software development life cycle (SDLC).

This issue was acknowledged by Jones and Rastogi \cite{jones2004secure} in 2004, who stated \textit{``The reality is that information security is an afterthought for many organizations''}.  Goertzel \cite{goertzel2008enhancing} also considered the problem, arguing that enhancing the SDLC would produce secure software.  Similar thoughts have also been voiced by Jerry Hoff - Vice President of the static code analysis division at WhiteHat Security in an article on DARKReading, stating the need to secure the SDLC \cite{Hoff2012}.  Hoff identifies that every organization will have their own approach to the SDLC and hence there is no one method on how to implement security throughout the process. However, he states that what is important is that once an organization has defined the security approach that work best for them, it is essential that they are consistent throughout development.

There are a variety of sources available providing guidance for developers implementing security into their SDLC, for example, SANS produced a document containing a variety of high-level suggestions on how to implement security at some of the key stages of the SDLC \cite{Haridas2007}.  Hasan \textit{et al.} \cite{hasan2017} have also stressed the importance of integrating the SDLC.  A model for ensuring security is built into the SDLC has previously been proposed \cite{futcher2007secsdm}, and others have investigated the use of such models, subsequently publishing a case study  \cite{karim2016practice}.

The objective of this approach is not to replace the need for alternative security testing - such as security audits and source code reviews, but rather to encourage developers to think about the security implications involved with the components they implement.

Furthermore, security researcher Peter Gregory stated that \textit{``Organizations that fail to involve information security in the life cycle will pay the price in the form of costly and disruptive events.''}\cite{Gregory2003}.  With developers now beginning to have an appreciation for the need for secure programming practises, it is possible that attack detection methods could be implemented into the application itself.

\subsection{AppSensor}
Although a number of commercial and open source Web Application Firewalls are available \cite{razzaq2013critical} \cite{singh2018impact}, there are alternative methods of enhancing web application, and increasing attack awareness. This research focuses on the approach of identifying malicious behavior from within the context of a web application.

In 2008 the Open Web Application Security Project (OWASP) began working on a project called AppSensor \cite{OWASP2017a}. AppSensor is an open source project originally developed by Michael Coates. The project was developed with the concept of being able to identify attacks from within the context of an application. If a developer is aware of the legitimate usage of their application, then they can confidently identify illegitimate and unintended usage. In 2017, OWASP stated that\textit{``The future of application defense is a system that can understand custom attacks against an application, correlate them against a malicious attacker, and react in real-time to contain and eliminate the threat"} \cite{OWASP2017a}.
Whilst the AppSensor project provides its own solution that can be adopted within a security environment, the main focus of this project is more to broadcast the idea of this type of attack detection. It has been stated that \textit{``the idea behind OWASP AppSensor project is that it is not a product, like a WAF would be, but rather an idea with a reference implementation in a Java framework"}  \cite{Thomassen2012}.

The objective of Thomassen’s work was to provide a comparison of the detection capability offered by web application firewalls, intrusion detection systems and AppSensor. Throughout the investigation, a variety of attacks based off of the top ten most common web application vulnerabilities \cite{OWASP2010} were used to determine the effectiveness of each solution.  The results demonstrated that the web application firewall used - modsecurity - provided satisfactory attack detection coverage, however, AppSensor proved a greater rate of attack detection.

With the concern for security growing throughout the cyber community, it was apparent there was a need for new security solutions and approaches to be introduced. In 2017 OWASP released an up-to-date list of the most common web application vulnerabilities, number ten on that list being insufficient logging and monitoring - \textit{``exploitation of insufficient logging and monitoring is the bedrock of nearly every major incident''} \cite{OWASP2017b}. With this issue being widely recognised within the industry, web application developers need to adopt more proactive methods for identifying attacks against their web applications. 

BlackWatch seeks to address this issue, exploring how to improve web application security by identifying malicious behaviour from within the context of a web application.  The following section will detail the methodology behind creating and evaluating the solution.


\section{Methodology}\label{methodology}

\subsection{AppSensor Investigation}
When considering the identification of malicious behaviour from within the context of a web application, OWASP's AppSensor was an influential solution in the field.  Prior to the development of the BlackWatch solution, AppSensor was analysed to gain an understanding of the strength and weaknesses of the tool.  Four areas were considered: 

\begin{itemize}
    \item Implementation- \textit{Is it feasible to implement AppSensor into an existing web application?}

    \item Configuration- \textit{Can AppSensor be easily configured to suit the requirements of individual web applications?}

    \item Attack Detection- \textit{Can AppSensor detect attacks whilst keeping
false positives and negatives to a minimum?}

    \item Reporting- \textit{Is the information gathered by AppSensor presented in an efficient manner?}
\end{itemize}

\subsubsection{Implementation}
As part of this research, AppSensor was implemented into MWR InfoSecurity’s internal application called ``The Bazaare''.  The Bazaare is an intentionally vulnerable ASP.NET web application used for training purposes within MWR.  This type of application was required for this research as it contains a large number of attack vectors which AppSensor could be configured to monitor.

Communications between The Bazaare and AppSensor used AppSensor’s Representational State Transfer (RESTful) API.  RESTful communication uses HTTP requests to communicate data between applications, requiring minimal changes to be made to the Bazaare environment.  AppSensor does not provide client-side libraries to aid in the implementation process, therefore a custom class was developed within the Bazaare that sends user events in the correct format using HTTP POST requests. One of the difficulties of this stage was generating the user ‘event’ in the correct format; the events being sent to AppSensor had to be in a specific format, including information such as: username, IP address, detection point etc.  Gaining an understanding of this structure and implementing this into an existing application highlights the challenges which developers face when attempting to increase attack awareness in their own web applications.

\subsubsection{Configuration}
The key attribute which makes AppSensor different to many other methods used to protect web applications, is its ability to work within the context of the web application itself. In order for AppSensor to achieve this contextual advantage it must be properly configured. This configuration stage was split into two main phases: identifying detection points and implementing response mechanisms. Identifying detection points is an essential step as it provides the areas where malicious activity can be identified. Implementing response mechanisms is what makes AppSensor effective at not only identifying malicious users, but also from stopping them exploiting the application.

\begin{figure}[ht]
\centering
\includegraphics[width=128mm]{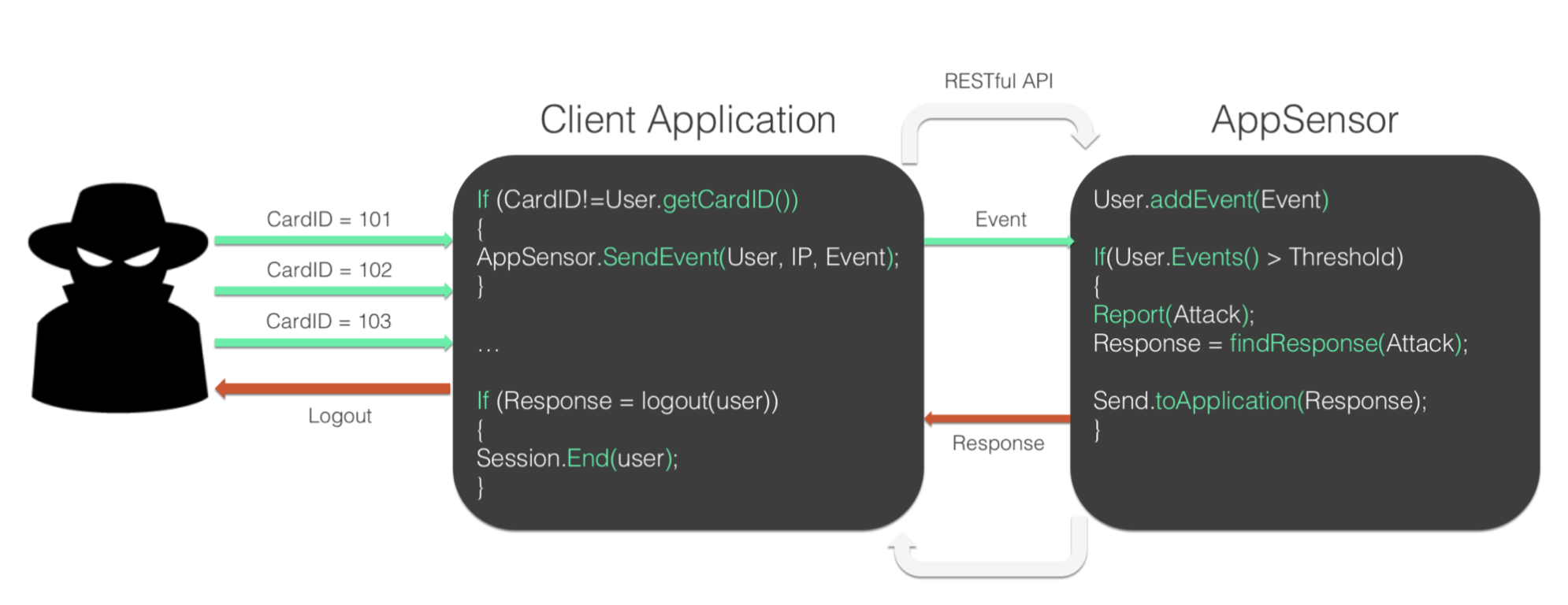}
\caption{AppSensor CardID example} 
\label{fig:appsensorcard}
\end{figure}

Figure \ref{fig:appsensorcard} demonstrates an example of how AppSensor works. Within this example the malicious attacker is altering the ‘CardID’ field of a POST request. If this vulnerability had not previously been remediated then this would allow the attacker the ability to use a CardID that does not belong to them. However, because a detection point has been implemented at this stage, the Bazaare application checks to see if the CardID submitted matches the user’s card.  If there is not match, the application discards the request and sends an event to AppSensor. AppSensor then uses information provided to determine if this activity exceeds the predetermined threshold that will identify this event as an attack. If an attack is identified, then AppSensor will alert the reporting mechanism and then provide the necessary response. In this scenario the response provided by AppSensor is to log the user out, however AppSensor allows for configuration of incremental responses e.g. if the same user triggered this attack again within a certain time frame, the response could be increased in severity, for example, disabling the user’s account.

\subsubsection{Attack Detection}
One of the main benefits of AppSensor is its attack detection capability is dependent on how the client application is configured. Providing developers with the ability to decide where malicious activity can occur within their application, allows for a high level of accuracy in detecting attacks.  AppSensor uses a rule based analysis mechanism to determine when user events have amounted to an attack.  Whilst this approach to attack detection can be very effective against basic attacks, it can be fairly easily bypassed by an experienced attacker that is aware of the security mechanisms in place.

\subsubsection{Reporting}
AppSensor provides a built in web reporting portal where application activity can be monitored.  This reporting mechanism displays information such as a live feed of events, attacks and responses as well as displaying the current configuration of AppSensor. Whilst this reporting mechanism is well presented, the underlying functionality is limited.

\subsection{ Monitoring Unauthenticated Users}\label{unauth}
Owing to the potential of attacks on web applications by malicious users, it is important to monitor unauthenticated users. Monitoring an authenticated user is simple as their actions can be associated with the account that they are performing certain activities under. Tracking the usage of users prior to authentication however is far more challenging endeavour. The
following section will discuss a variety of methods that have been investigated in regards to monitoring these unauthenticated users.

\subsubsection{IP Addresses:} One of the most common methods used for monitoring individual users throughout traditional monitoring tools - such as intrusion detection systems - is by monitoring the user's IP address. Monitoring IP addresses can be very effective in some circumstances, especially since IP addresses can be easily blocked at the networking layer. However, there are two main disadvantages of monitoring using this information: IP addresses are shared and they can be easily changed. As a result of how IP address allocation works, an IP address does not necessarily only belong to one user, but may rather belong to a large group of users within the same network. As a result, blocking IP addresses may affect legitimate user functionality, as well as this, it is a trivial step for an experienced attacker to alter their IP address using
a tool such as a virtual private network (VPN) \cite{Belesi_2016}. 

\subsubsection{Session Variables}\label{sessionvar}
Most web applications use session variables such as cookies or session identifiers to track users throughout the application. This method is one of the most widely adopted mechanisms for tracking application usage, as it can be very effective for monitoring individuals users due to the fact each session variable is unique.
However, the issue that accompanies web applications using this method, is that for this monitoring mechanism to work, users must willingly submit these session variables to the server. As a result an attacker could simply strip these session variables from any communication they have with the server, and the web application
would simply generate the user new session variables with each communication.

\subsubsection{Device Fingerprinting}
This information may include details such as: the client's operating system, the browser being used, plugins installed within the browser and the screen resolution of the device. This type of information is often gathered by web applications to provide a more tailored experience for the user, such as targeted marketing. However, this information may also be useful for identifying potentially malicious users. Whilst many users will have similar fingerprints and hence be near impossible to distinguish using this approach, certain attributes may cause malicious users to `standout'. For example the majority of users are most likely going to be using common
operating systems such as: Windows, Mac OS, Android and iOS. This means that if a user device is running an operating system such as Kali Linux (an operating system designed for penetration testers) then this user may be identified as `suspicious' and therefore their actions should be monitored closely \cite{abouollo2017detecting}. 

Factors outlined in this section must be considered when designing a solution which aims to increase attack awareness.
 
\subsection{Design}
A solution was developed that allows applications the ability to identify attacks from within the context of the application; and to respond effectively to deter attackers. The development process of this solution was influenced by two  important factors: application scalability and well documented and structured code.

\begin{figure}[ht]
\centering
\includegraphics[width=108mm]{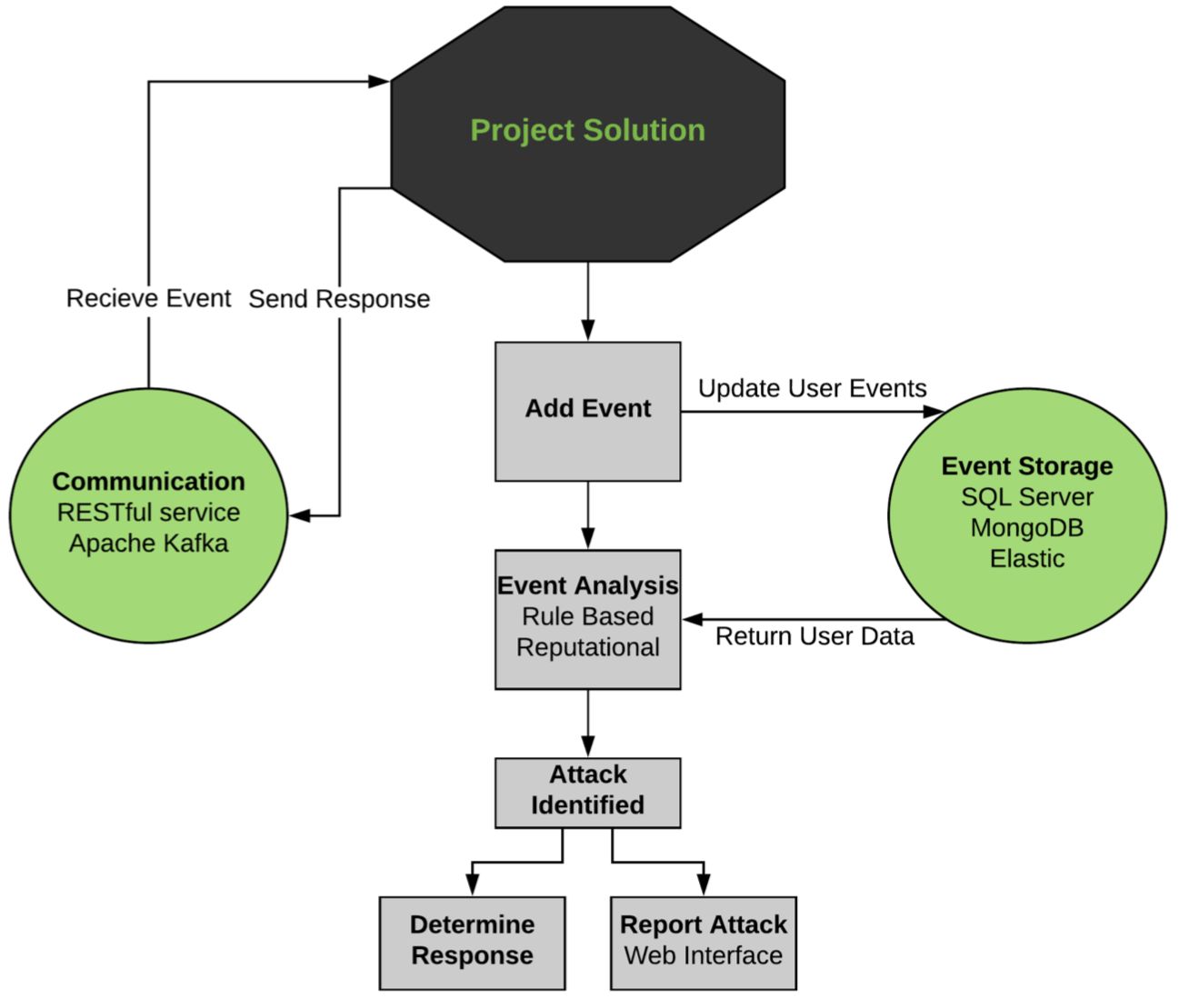}
\caption{BlackWatch system overview}
\label{fig:blackwatchoverview}
\end{figure}

One of the first stages of the development process was to design the structure of the application. The structure diagram shown in Figure \ref{fig:blackwatchoverview}  demonstrates the initial design of this project’s solution - hereby named ‘BlackWatch’. 

\subsection{Implementation}
\begin{figure}[ht]
\centering
\includegraphics[width=98mm]{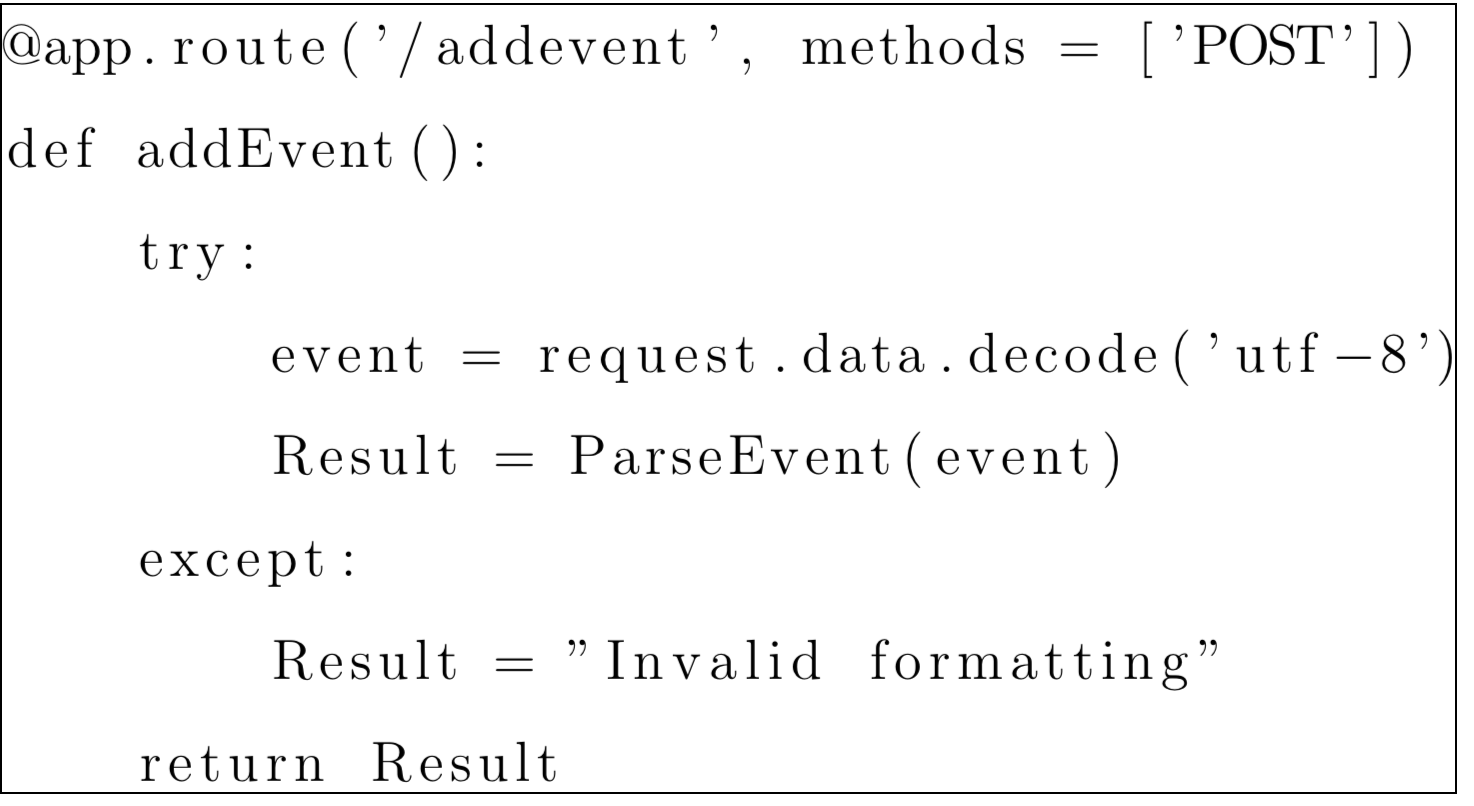}
\caption{RESTful server- POST method}
\label{fig:restfulpost}
\end{figure}

BlackWatch was implemented in the Python programming language. One of the first stages of the development process was to design the structure of the application. The first step taken within the implementation stage was to develop a RESTful API service that would allow for communication between the client application and the BlackWatch solution. A RESTful server uses HTTP requests for communicating with clients, the HTTP protocol uses a number of different methods for sending and receiving data; for the BlackWatch solution the methods used are POST and GET requests. Using flask, this type of functionality can be easily implemented as shown in Figure \ref{fig:restfulpost}, where a POST request will be received and then processed in attempt to retrieve information regarding an application event.

The remaining components that had to be implemented into the BlackWatch application were split into four general areas; handling events, identifying attacks, determining responses and reporting. 

\subsubsection{Handling events}
During the first stage, handling events, it is essential to ensure that the event received from the client application contains all of the necessary information in the correct format. Each event must contain two main objects: ‘User’ and ‘Detection Point’, the first object contains information identifying the user responsible for the event, and the second where the event was triggered. This information must be present - and correctly formatted - in order for the event to proceed further within the application. Given that this information is correct, the event must then be added to the BlackWatch database and sent for analysis.

BlackWatch uses the MongoDB database solution \cite{MongoDB2018}. PyMongo allows the BlackWatch application the ability to add and retrieve information from the database with ease, as well as including functionality for performing filtered queries to retrieve relevant information from the database quickly. The documents stored within MongoDB are done so in a Binary JSON format (BSON); very similar to the JSON format used throughout the BlackWatch application. 

\subsubsection{Identifying attacks}
Once an event has been added to the BlackWatch database, the next step is to send this event through the analysis process. The purpose of the analysis process is to identify whether an event holds enough merit to be identified as an attack. There are two main analysis mechanisms within the BlackWatch solution used to determine this; rule based and reputation based analysis. Rule based analysis is a very simple method of attack identification that has been used within security mechanisms for years. The idea being that if a user triggers a predetermined number of events within a specific area during a certain period of time, then these events are identified as an attack and a response is set accordingly. For example, if the user ‘Bob’ triggers an event at the ‘Login Page’, then this does not necessarily mean that Bob is performing malicious activity, it may simply be a mistake. However, if Bob were to trigger this event ten times within thirty seconds, then it is highly likely that this is an attack and hence the application should respond. This type of attack detection can be very effective; however, its main downfall is that it does not take into consideration a user’s activity throughout other areas of the application.

When rule based analysis does not identify an event as an attack, the reputation based analysis mechanism is invoked. The reputation based analysis mechanism analyses a users behaviour over the entire application within a certain time frame. The idea being that whilst a user has not triggered enough events in one specific area to trigger an attack, they may have shown a pattern of malicious activity across the application. In order for this mechanism to work, each detection point is given a level of severity. For example a user inserting special characters into a search function could very easily be a mistake, hence this detection point will be given a ‘Very Low’ severity rating. However, a user uploading a malicious file to an application is almost certainly an attack and hence should be given a ‘High’ severity rating. These severity ratings hold a certain weighting within the BlackWatch application as shown in Table \ref{table:ratings}.

\begin{table}[ht]
\centering
\begin{tabular}{|p{6cm}|p{3.5cm}|}
\hline
\rowcolor{ashgrey}
\textbf{Severity Rating} & \textbf{Value} \\ \hline
 High & 8 \\ \hline
 Medium & 4  \\ \hline
 Low & 2 \\ \hline
 Very Low & 1  \\ \hline
\end{tabular}
\caption{Severity ratings and values}\label{table:ratings}
\end{table}

The values of the severity weighting increase by powers of two, representing an incremental structure whereby high must have a greater value than low.  Every time a user triggers an event, the severity weighting belonging to the event’s detection point will be added to the user’s reputation. Once a user’s reputation exceeds nine, the user is identified as malicious and an attack is triggered. One of the most important features of the reputation based analysis mechanism is that a user’s reputation must decrease over time; this ensures that if an attacker stops abusing the application, their reputation must reflect their legitimate behaviour. To achieve this feature, the application will decrease the users reputation by one every ten minutes until the reputation is equal to zero.

One of the main advantages of each of the mechanisms discussed above, is that both have the ability to monitor unauthenticated users. Both mechanisms will check to see if a username has been provided for the user in question, and if not will handle the analysis based on the user’s session ID. This approach allows BlackWatch the ability to monitor unauthenticated users throughout the application, and if necessary to respond to their malicious activity.

\subsubsection{Determining responses}
Upon successful identification of an attack, the necessary response must be determined. The BlackWatch application allows users the ability to set multiple responses for the same detection point. This allows for the responses chosen to be done so in an incremental manner; meaning that if a user has triggered an attack at this same detection point within the past thirty minutes, then the response chosen can be incremented to take more severe action. Once the necessary response has been identified, it is then added to the ‘Responses’ MongoDB collection. Client applications can access these responses using an HTTP GET request containing the intended username and session ID.

The BlackWatch application will return any responses specific to that user and then delete these responses from the database. This is to prevent any duplicate responses from being returned to the client application. The reason that responses are sent to the client application in this manner and not simply returned in response to the initial event sent from the client, is that this approach ensures that the client application’s performance is not affected whilst the BlackWatch application analyses the event. Once responses are sent to the client application, it is then up to that application to execute these responses using its own functionality. The main benefit that arises from responses being executed in this manner, is that the only limitations to what responses can be set, are bound by the functionality of the application itself.

At this stage of the development process the BlackWatch application contains the ability to: receive suspicious events from client applications, analyse these events to determine if they indicate an attack, and if so identify the necessary response. Whilst this functionality is what makes the BlackWatch solution effective at deterring malicious attackers, the overall objective of this project is to raise attack awareness within web applications and hence this functionality must provide a well structured reporting mechanism.

\subsubsection{Reporting}
Rather than simply reporting BlackWatch’s activity using log files like so many other security solutions, this application uses a web interface to display the relevant information in a professional and efficient manner. One of the main challenges that faced this stage was the decision on how to communicate between the web application client (the reporting mechanism) and the web server (the BlackWatch application). In order for the reporting mechanism to reflect the live activity that is happening within the BlackWatch application, it was essential for the web server to be able to send events directly to the reporting application. However, there lies a difficulty in that traditionally web applications only communicate with the server when they need to, and hence the server has no way of sending information unless it is first requested \cite{West2018}. 

\begin{figure}[ht]
\centering
\includegraphics[width=118mm]{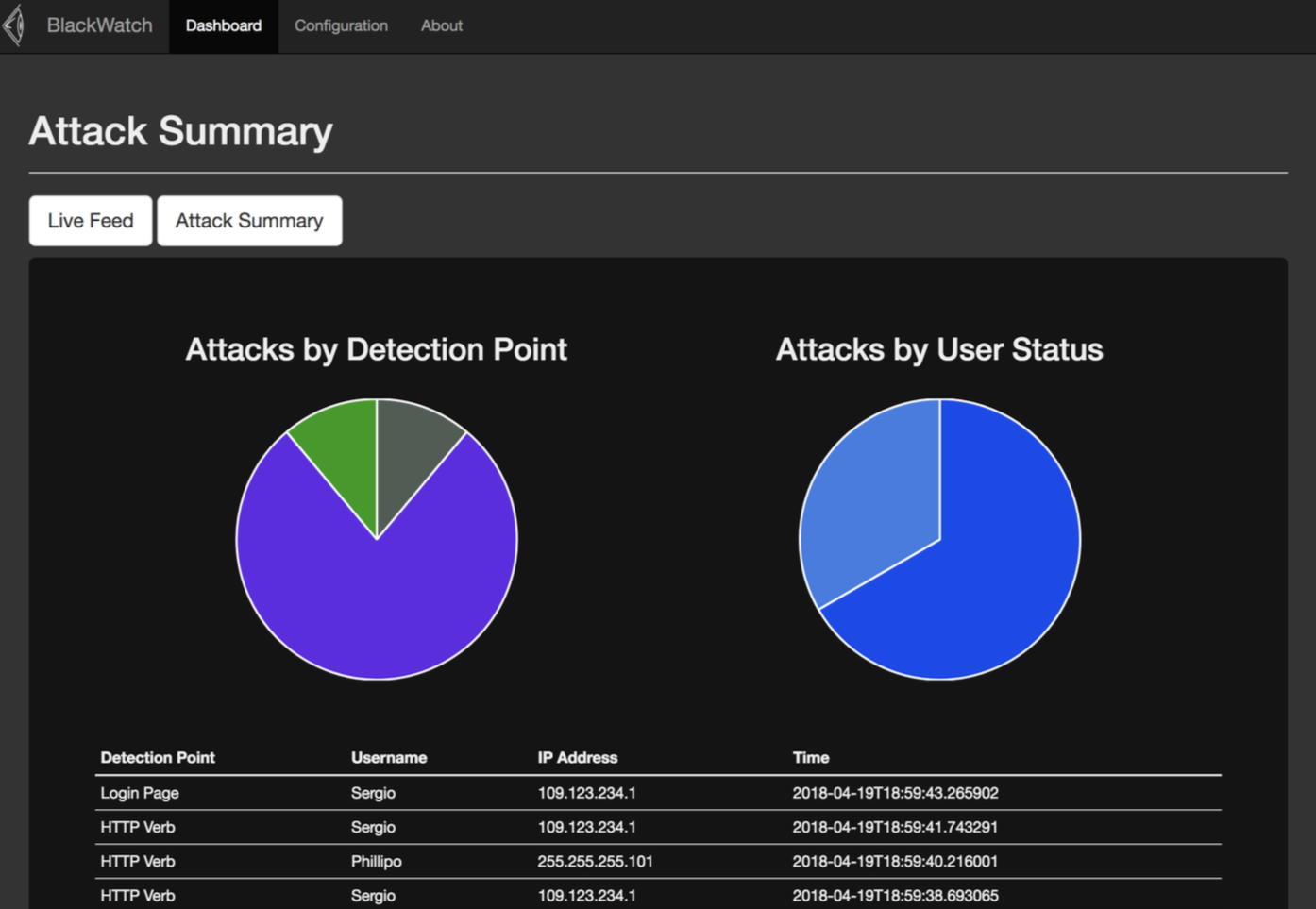}
\caption{Attack summary}
\label{fig:attacksummary}
\end{figure}

To tackle this issue WebSocket technology was implemented. WebSockets are a method used for creating bidirectional communication between client-side applications and web servers. JavaScript has a standard library called Socket.IO which is used within the web reporting application. The Flask library Flask-SocketIO is used within the BlackWatch application to allow the server to implement WebSocket technology, and to send information to the web application on demand \cite{Miguel2018}.

As well as displaying the activity that is taking place within the BlackWatch application (as demonstrated in Figure \ref{fig:attacksummary}), the reporting mechanism can also be used for configuring detection points to be used within your application. The ability to create and remove detection points from within the web reporting mechanism makes it easy for users to configure the BlackWatch solution to fit the needs of their web application.

Following the development of the BlackWatch solution, an evaluation was performed. The success of BlackWatch is based on the coverage of four main areas: ease of implementation, ability to identify unauthenticated users, accuracy of attack detection, effectiveness and efficiency of reporting.  The subsequent section highlights the results, before providing an analysis in section \ref{discussion}.

\section{Results}\label{results}

\subsection{Unit Tests}
The unit tests created during the design stage of this project were developed to evaluate the BlackWatch application’s ability to: receive events, detect attacks using both rule based and reputation-based analysis and also to retrieve responses from the application. 

The BlackWatch application passed all unit tests with the exception of the response retrieval test; this test was designed to trigger an attack and to then try and retrieve the correct response from the BlackWatch application. During the development process the approach used to communicate responses changed, therefore causing this unit test to fail. Despite failing this unit test, the overall results from these tests were positive, demonstrating that BlackWatch could not only detect attacks by authenticated users, but also by monitoring unauthenticated users through the use of session identifiers as discussed in section \ref{unauth}. Table \ref{table:unittests} demonstrates the results of the unit testing phase.

\begin{table}[ht]
\centering
\begin{tabular}{|p{4cm}|p{3.5cm}|p{3.5cm}|p{1cm}|}
\hline
\rowcolor{ashgrey}
\textbf{Unit Test} & \textbf{Result}  & \textbf{Intended Result} & \textbf{Pass or Fail} \\[20pt] \hline
\rowcolor{grannysmithapple}
 Send Correctly Formatted Event & ``Event is being added'' & ``Event is being added'' & PASS  \\[20pt] \hline
 \rowcolor{grannysmithapple}
Send Invalid IP Address within Event & ``Incorrect formatting within POST request'' & ``Incorrect formatting within POST request'' & PASS  \\[20pt] \hline
\rowcolor{grannysmithapple}
 Detect Attack using Username & Attack shown within the reporting mechanism & Attack shown within the reporting mechanism & PASS  \\[20pt] \hline
 \rowcolor{grannysmithapple}
 Detect Attack using Session ID & Attack shown within the reporting mechanism & Attack shown within the reporting mechanism & PASS  \\[20pt] \hline
 \rowcolor{grannysmithapple}
 Detect Attack using Reputation Based Analysis & Attack shown within the reporting mechanism & Attack shown within the reporting mechanism & PASS  \\[20pt] \hline
 \rowcolor{lightsalmonpink}
 Retrieve Response & ``Failed'' & ``Response Returned'' & FAIL \\[20pt] \hline
\end{tabular}
\caption{BlackWatch unit test results}\label{table:unittests}
\end{table}

\subsection{User-Based Evaluation}
A total of five participants took part in stage of the evaluation. The number of participants may seem relatively small however the evaluation process took over an hour to complete, and individuals were participated on a voluntary basis.  Therefore, due to time constraints it was only feasible to include a small sample size during this stage.  

The results gathered from the user-based evaluation process are split into four main sections: participant skill- set, BlackWatch setup and configuration, attack detection and reporting. These results consist of both quantitative and qualitative data. 

\subsubsection{Participant Skillset} \label{skillset}
The evaluation process considered the development skillset possessed by participants.  Participants were asked to rate their overall level of development experience on a scale of 1 to 5 (whereby 5 equalled Very Experienced) as part of a questionnaire.  Overall, results showed participants considered themselves to have an average level of development experience (\textit{n=5, mean=3.2}).

Participants were also asked to rate their development knowledge with both PHP, and Python.  Overall results showed that whilst participants considered themselves to have an average level of development knowledge in Python (\textit{n=5, mean=3.2, mode=3}), they generally felt they had a lower level of knowledge in PHP (\textit{n=5, mean=1.6, mode=1}).

\subsubsection{BlackWatch Setup and Implementation}
The questions within this section aimed to identify participants experience during the setup and implementation of the BlackWatch solution. The practical steps required during this stage were guided by a tutorial provided by the BlackWatch solution; all participants stated that these instructions covered all of the necessary areas throughout the setup procedure. 80\% of participants rated both the BlackWatch setup and the client side implementation as being `very easy'; the 20\% rated it as `easy' (\textit{n=5, mean=4.8}, where 5 equalled very easy).
None of the participants recorded any difficulties faced throughout this stage and no recommendations were made in regards to improving the instructions provided.

\subsubsection{Attack Detection and Reporting}
Upon taking on the role of a malicious attacker targeting the web application, participants rated the BlackWatch applications accuracy of attack detection as having a mean score of 4.8 - one being not accurate, five being very accurate - meaning that the malicious activity they were performing was being detected. A mean score of 4.8 was also given to the effectiveness of the responses provided by the BlackWatch application - one being not effective, five being very effective. Within this section participants were asked whether or not they believe that the BlackWatch application would ``deter attackers from targeting that application'', every participant responded `Yes'. Whilst the most important aspect of this project was the BlackWatch application’s ability to detect attacks, it was also essential that the application provided a well structured reporting mechanism. Participants were asked to rate their experience using the web reporting mechanism, 80\% of these participants agreed that the mechanism provided was very efficient with one participant rating the mechanism’s efficiency as 4 out of 5 - one being not efficient, five being very efficient (\textit{n=5, mean=4.8}).

\subsubsection{Overall}
One of the key questions asked throughout this evaluation process was whether or not participants believed that, if properly implemented, the BlackWatch solution would increase attack awareness within web applications. Every participant that took place within this evaluation agreed `Yes'. The final section of the survey asked participants for any additional comments that they have relating to the BlackWatch application; the majority of comments given during this stage praised the design and usability of the reporting mechanism used by the BlackWatch application.

\section{Discussion}\label{discussion}

\subsection{Ease of Use}
It is vital for the BlackWatch solution to be considered user-friendly.   In order for this type of solution to attract developers and security professionals, it must be easy for them to understand and implement. Setting up this solution proved to be simple, even for participants that rated their development experience at the lower end of the experience scale still managed to setup and implement the BlackWatch solution with ease.

Interestingly the results demonstrated in section \ref{skillset} indicate that the participants who took place within the evaluation process felt far more knowledgeable of the Python programming language than they did with PHP. This was one of the main reasons that this programming language was chosen, as Python’s easy to learn syntax ensures that developers interested in using this solution, can review the underlying code in attempt to understand the full functionality of the BlackWatch solution.

\begin{figure}[ht]
\centering
\includegraphics[width=128mm]{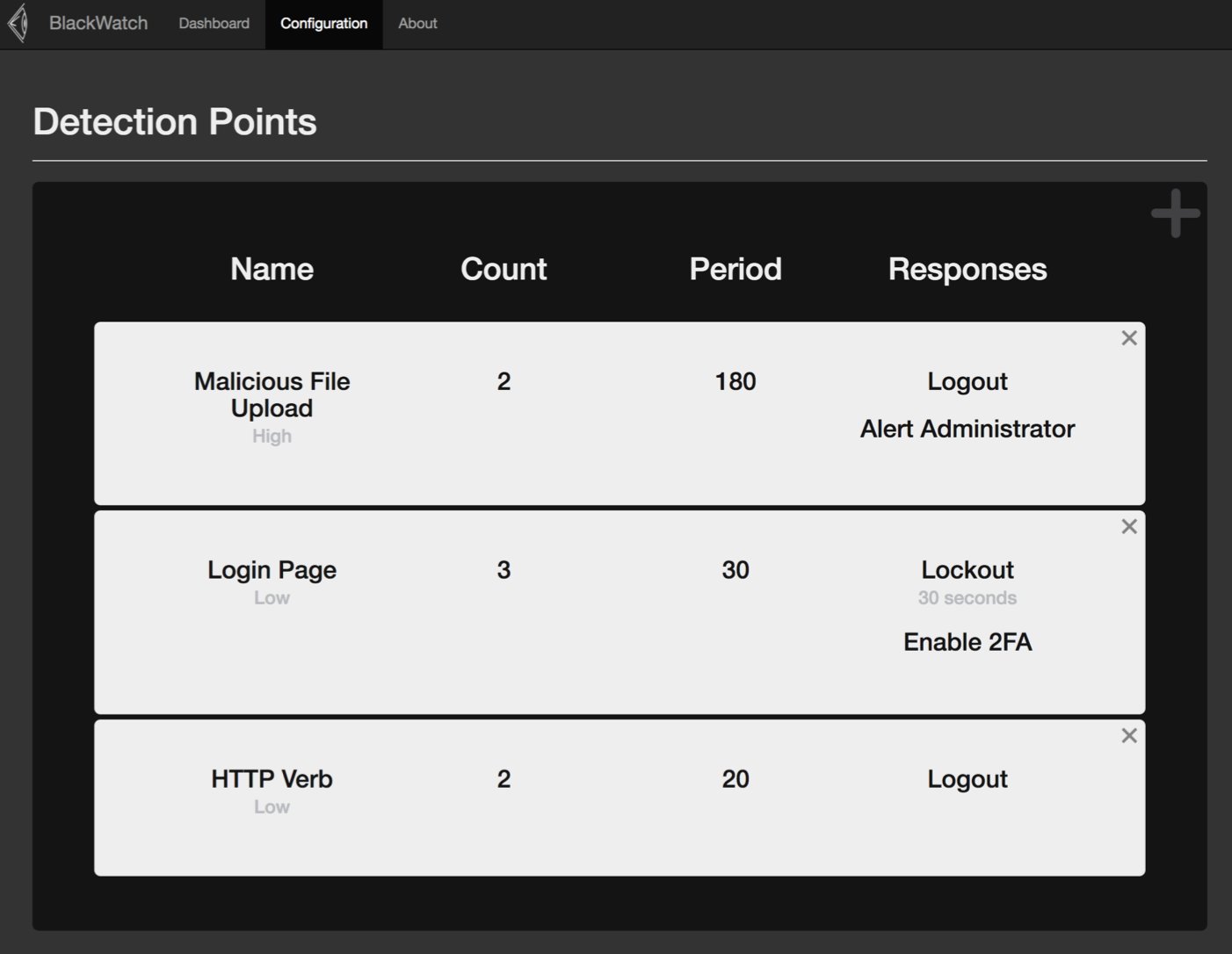}
\caption{Configured detection points listed in BlackWatch} 
\label{fig:detectionpoints}
\end{figure}

The area that requires the most effort from those implementing this project’s solution, is the configuration of detection points within the client application. Often security solutions are based outwith the applications they protect. However, the attribute that makes the BlackWatch solution effective at identifying attacks, is that it works within the context of the application. Developers have a clear understanding of the legitimate usage of their applications and hence they should be able to accurately determine what classifies as malicious, or suspicious behaviour. Therefore, in order to implement the BlackWatch solution into their application (Figure \ref{fig:detectionpoints}), developers simply need to identify points where they can detect this malicious behaviour and insert the necessary code to communicate with the BlackWatch solution. 

The evaluation of this project made use of a pre-written PHP library that allowed users to very easily communicate with the BlackWatch solution, this client-side library could be easily replicated into other programming languages as discussed within this project’s future work. During the evaluation procedure participants found the creation of detection points to be easy, even though they were not familiar with the client application they were working with. This indicates that the implementation process should be very easy for a developer, as they will have prior knowledge of the web application environment they are working within.

One concern that surrounds this approach of attack detection within web applications, is that in order for a developer to know where to implement detection points within their application, they must be aware of the type of attacks a malicious user may perform.

Whilst this may be an easy task for developers with security experience, it may pose as a difficulty for those who have never performed any research into the offensive side of security. A possible action that could be taken to counteract this issue, would be to involve a security professional within the implementation stage of the BlackWatch solution. A security professional, such as a penetration tester, would be able to advise on the areas that an attacker is likely to attack, and how these attacks could be identified with a high degree of accuracy.

\subsection{Attack Detection}
OWASP \cite{OWASP2017b} states that \textit{``exploitation of insufficient logging and monitoring is the bedrock of nearly every major incident''}. As discussed previously, a lack of attack detection can leave an organization vulnerable to advanced persistent threats. During the evaluation of BlackWatch, participants were asked to portray the role of a malicious attacker attempting to exploit the Damn Vulnerable Web Application (DVWA).

BlackWatch monitored participants’ activity in an attempt to identify attacks and respond accordingly. The results from this section indicate that the malicious activity performed by each participant was being detected to a high degree of accuracy. 

\subsubsection{False Positives and Negatives}
When developing a security product one of the main concerns that must be addressed is the occurrence of false positives and negatives. False positives occur when legitimate behaviour is wrongly identified as an attack. False negatives are the exact opposite - when an attack is wrongly identified as legitimate behaviour and hence no action is taken to reprimand the attacker. False positives are often the more severe issue for organizations as they can affect legitimate users and negatively impact user experience. 

During the evaluation of this solution, false positives were not found to be an issue as the detection points created within the client web application took user error into consideration. For example, within the client application used there was a component that could be used to ‘ping’ an IP address. Functionality like this would often be tested by malicious users for vulnerabilities such as command injection \cite{OWASP2016}. Hence, a detection point was configured here to detect any input that does not match the application’s expectations. As an attempt to reduce false positives, if the user input matches the correct format of an IP address, but is not a legitimate address eg. ‘455.455.1.1’ (each octet must be below 256) then no event is sent to the BlackWatch application, as this could have been a user error.

The reputation-based analysis mechanism within the BlackWatch application is used as an attempt to reduce false negatives, as it attempts to correlate events that are being triggered throughout the application by individual users. The unit testing performed during the evaluation of the BlackWatch solution demonstrated the application’s ability to identify attacks based on a pattern of events occurring throughout different components of the client application, rather than the abuse of one single detection point. This detection technique can help identify attackers who may be attempting to avoid the rule-based detection mechanism. However, this approach also introduces the risk of producing false positives. With rule-based detection it is easy to identify an attack based on the repetition of one suspicious event. However, with reputation-based analysis it is more likely that a legitimate user could trigger a number of different ‘suspicious’ events throughout multiple areas of the application through user error. Despite this increased risk, none of the participants within the evaluation process experienced any false positive or false negative errors.

\subsubsection{Monitoring Unauthenticated Users}
It was essential for this project that there was some mechanism in place for monitoring unauthenticated users and their actions. Discussed in section \ref{unauth} are numerous different approaches that can be used in an attempt to monitor users who have yet to authenticate within the client application. 

One of the most common approaches security solutions take to monitor these users is by tracking their IP address, however, due to this approach’s lack of accuracy - and potential implications to legitimate users - another approach was taken instead. 

Session variables such as session IDs are provided to all users of an application as a means of ensuring that the content they see is specific to their individual session. The results from the evaluation stage demonstrate that this session variable can be used as a method of user monitoring, and can effectively associate attacks with individual users. However, despite the effectiveness of this method against novice attackers, an experienced attacker would most likely be able to figure out that their session ID is being used to monitor their attacks, and hence they could quite easily remove this information from the requests they send to the web application. Even though it is possible for an attacker to bypass this method of activity monitoring, it is still an effective approach that adds an extra layer of security to the web application.

\subsubsection{Reporting}
\begin{figure}[ht]
\centering
\includegraphics[width=128mm]{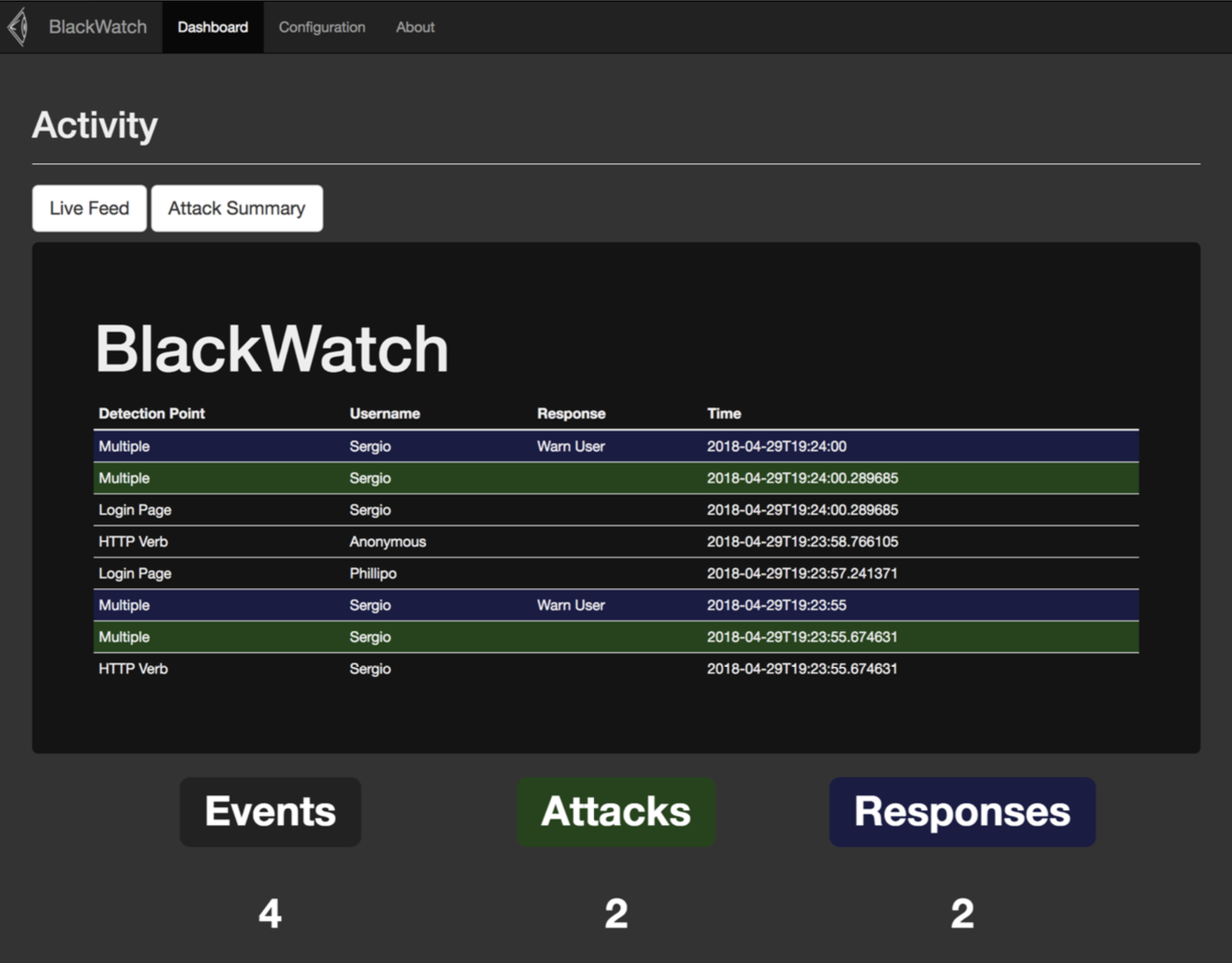}
\caption{BlackWatch dashboard}
\label{fig:dashboard}
\end{figure}

The reporting mechanism provided with the BlackWatch application allows administrators the ability to monitor the events and attacks occurring within their application (Figure \ref{fig:dashboard}). According to the results from the evaluation stage, participants found the reporting mechanism to be very well designed and user friendly. Throughout this project the reporting mechanism developed was created with the intention of being simple and efficient, however, given further development there are many features that could be added to improve this aspect of the BlackWatch application.

\subsection{Attack Response}
Having the ability to accurately detect malicious users within an application gives developers the ability to respond with deterring action. Giving applications the ability to detect attacks was the main objective of this research.  It was also important to investigate methods that could be used to respond to these attacks. During the evaluation of this project the main responses used were:

\begin{itemize}
    \item User Warning
    \item User Logout
    \item Page Redirect
    \item Fake Output (such as mock database) information
\end{itemize}

The `fake output' was a response used at a detection point aimed to identify SQL injection attacks against the client web application being tested. The purpose of this response was to provide false information to the attacker in an attempt to confuse, and or, waste their time. Participants involved during the evaluation stage all stated that they believe these responses would successfully deter or slow down an attacker. As stated within the methodology stage of this project, one of the key aspects of how the BlackWatch application operates, is that the responses that can be used are only limited by the client web application's functionality.

\subsection{Challenges}
During the initial research performed into this project area it was apparent that this approach to attack detection was still very much a concept that had yet to be exposed to the security landscape. This introduced a number of difficulties as it was often difficult to find the relevant literature to help guide the development.

The most challenging task involved with this approach of attack detection is being able to encourage developers to take the extra step of implementing this functionality into their application. Whilst the results from the evaluation stage indicate that the BlackWatch solution is easy to setup and implement, it is still an extra step that a developer would need to take within the already complex process of developing an application. However, as discussed in section \ref{sdlc}  implementing security characteristics such as attack awareness into an application during the development process can help prevent security incidents and hence save organizations both time and resources in the long run.

\section{Conclusions}\label{conclusions}
This research demonstrates that it is no longer enough for organizations to focus their efforts towards preventing attacks, but rather stresses the importance of detecting the presence of these attacks in the first place. Being able to accurately detect attacks within a web application can allow organizations the ability to execute responses with the objective of detecting, or slowing down attackers. 

The methods currently used for detecting attacks within the cyber community often lack the necessary configuration required to produce effective detection rates and hence attacks often go unnoticed. 

The BlackWatch application developed throughout this project attempts to increase the levels of attack awareness by working within the context of web applications themselves. By implementing points within applications where malicious activity can be detected, developers are given the ability to confidently identify user behaviour that does not appear as legitimate usage. This behaviour is then correlated outwith the web application itself in order to determine if a user has demonstrated a pattern of malicious behaviour. If so an attack is identified, and the appropriate response is determined.

A concern that accompanied this project and its approach to attack detection, was the challenge of encouraging developers to implement this type of security into their application. To ensure this issue did not hinder the research objective of increasing attack awareness, the BlackWatch application was developed with ease of use being one of the key aspects. Through a preliminary user-based evaluation it was demonstrated that not only is the BlackWatch application easy to setup, but also it is also easy to configure within a client environment. One of the factors that contributes to this ease of use is that developers are implementing the solution into their application - a programming environment they fully understand. It is this usability factor combined with the BlackWatch solutions' accuracy of attack detection, that highlights the success of this research in providing a method of increasing attack awareness within web applications.

\subsection{Future Work}
When developing the BlackWatch solution, the objective was to create a proof of concept.  However, it is essential that upon further testing and research into this project area that new analysis mechanisms are introduced to continually improve the accuracy of attack detection. The web reporting mechanism used by the BlackWatch application could also be improved by implementing more advanced components e.g., during the design stage of this project the idea of manual responses was considered; manual responses would allow an administrator to manually set the responses for malicious users from within the reporting mechanism. 

To ensure that the BlackWatch application can be widely used throughout a variety of web applications, it is essential that more client libraries are developed to allow the BlackWatch solution to be easily implemented into the most common web application frameworks.

An area that was not covered during this research, was the deployment of the BlackWatch solution within a production environment. The objective of this work was to demonstrate the level of effectiveness provided by this method of attack detection. However, the solution was only evaluated within a test environment, and hence its performance has never been tested within a production scenario. Prior to deploying the BlackWatch application alongside a real-world application, developers must firstly investigate a number of different areas to ensure that the BlackWatch application performs in an efficient, effective and reliable manner. One of the main areas that must be researched is the server used to host the BlackWatch application. The server used to host the BlackWatch application will be one of the main factors that determine how well the application performs, and hence it is crucial that future work involving server selection is carried out to help guide developers.

\vspace{6pt} 



\authorcontributions{C.H. carried out the investigation, and wrote the first draft of the paper. L.A.S., and
N.C. supervised the project, and wrote the paper.}

\funding{This research received no external funding.}


\conflictsofinterest{The authors declare no conflict of interest.} 



\reftitle{References}
\externalbibliography{yes}
\bibliography{references}



\end{document}